\documentclass[prd,twocolumn,nofootinbib,reprint,superscriptaddress]{revtex4-1}
\usepackage{amsmath,amssymb,comment,url,hyperref,graphicx,color,fancybox,slashed}
\begin{document}
\rightline{OU-HET-1132}
\title{Chiral Symmetry Breaking and Quark Bilinear Condensate in Large N QCD}
\author{Ryosuke Sato}
\affiliation{Department of Physics, Osaka University, Toyonaka, Osaka 560-0043, Japan}
\email{rsato@het.phys.sci.osaka-u.ac.jp}
\date{\today}
\begin{abstract}
\vspace{1mm}
We discuss spontaneous chiral symmetry breaking and quark bilinear condensate in large $N_c$ QCD.
It has been known that the existence of $\eta'$ meson is implied in large $N_c$ QCD as pointed out by Witten and Veneziano.
First, we show that the existence of $\eta'$ and Ward-Takahashi identities Eqs.~(\ref{eq:WT singlet}, \ref{eq:WT adjoint}) imply the existence of the Nambu-Goldstone bosons from chiral symmetry breaking $SU(N_f)_L \times SU(N_f)_R \to SU(N_f)_V$.
Second, we show that a QCD inequality Eq.~(\ref{eq:weingarten inequality}) implies a nonzero lower bound Eq.~(\ref{eq:bound on qqbar}) 	on the quark bilinear condensate.
\end{abstract}
\maketitle

\section{Introduction}
The chiral symmetry breaking \cite{Nambu:1961tp, Nambu:1961fr, Goldstone:1961eq} is one of the most important features in quantum chromodynamics (QCD).
QCD with $N_f$ flavors of massless quarks has continuous global symmetry $SU(N_f)_L \times SU(N_f)_R \times U(1)_V \times U(1)_A$ in the classical Lagrangian. 
$U(1)_A$ symmetry is explicitly broken by quantum anomaly \cite{Adler:1969gk, Bell:1969ts, Belavin:1975fg, tHooft:1976rip, tHooft:1976snw},
and the global symmetry at quantum level is $SU(N_f)_L \times SU(N_f)_R \times U(1)_V$.
It is shown that $SU(N_f)_V \times U(1)_V$ is unbroken in the vacuum \cite{Vafa:1983tf}.
On the other hand, it is believed that the chiral symmetry is spontaneously broken as $SU(N_f)_L \times SU(N_f)_R \to SU(N_f)_V$ if $N_f$ is below some threshold value.
Although the chiral symmetry breaking is successful to describe hadron physics, we do not understand why the chiral symmetry is broken yet.
There are numerical evidences of chiral symmetry breaking in lattice QCD calculation (see, \textit{e.g.}, Ref.~\cite{Faber:2017alm} and the reference therein).
Also, there are investigations from anomaly matching \cite{Frishman:1980dq, Coleman:1982yg},
supersymmetric QCD \cite{Aharony:1995zh, Alvarez-Gaume:1996vlf, Alvarez-Gaume:1997wnu, Martin:1998yr, Cheng:1998xg, Strassler:1997ny, Kitano:2011zk, Murayama:2021xfj},
and holographic QCD \cite{Babington:2003vm, Kruczenski:2003uq, Sakai:2004cn, Erlich:2005qh, DaRold:2005mxj}.

In this paper, we discuss the spontaneous chiral symmetry breaking
in QCD-like theory, \textit{i.e.}, $SU(N_c)$ gauge theory with $N_f$ flavors of quarks in the limit of large $N_c$ \cite{tHooft:1973alw}.
First, in Sec.~\ref{sec:massless}, we review Witten-Veneziano relation \cite{Witten:1979vv, Veneziano:1979ec},
which claims the existence of a light particle whose mass scales as $1/\sqrt{N_c}$,
and this particle can be interpreted as $\eta'$, (pseudo) Nambu-Goldstone (NG) boson from spontaneous breaking of $U(1)_A$ symmetry.
Then, in Sec.~\ref{sec:massive}, we will see that consistency with the existence of $\eta'$ and Ward-Takahashi identities with quark mass
implies the existence of light scalar particles which are associated with axial current.
Those particles become massless in the massless quark limit, and they are nothing but Nambu-Goldstone (NG) bosons from chiral symmetry breaking.
Finally, in Sec.~\ref{sec:qqbar}, we estimate a lower bound on $\langle \bar q q \rangle$ by a QCD inequality \cite{Weingarten:1983uj} in the same way of Kogan-Kovner-Shifman \cite{Kogan:1998zc}, and see $\langle \bar q q \rangle$ becomes nonzero.

%
%

\subsection{Comparison with the previous literature}
Let us compare our discussion with the previous literature.
Proofs of chiral symmetry breaking in large $N_c$ QCD by Coleman-Witten \cite{Coleman:1980mx} and Veneziano \cite{Veneziano:1980xs} have been known.
Coleman-Witten \cite{Coleman:1980mx} has shown the existence of massless scalar poles which are associated with axial current by anomaly matching in a elegant way, however, they have simply assumed the order parameter is bi-fundamental of $SU(N_f)_L \times SU(N_f)_R$ and have not concluded $\langle \bar q q \rangle \neq 0$.
Veneziano \cite{Veneziano:1980xs} uses the Ward-Takahashi identities which are equivalent to our Eqs.~(\ref{eq:WT singlet}, \ref{eq:WT adjoint}). They claims $\langle \bar q q \rangle \neq 0$, however, they have not realized an subtle issue which will be discussed in Sec.~\ref{sec:qqbar nonzero} and footnote \ref{footnote:stern phase}.
In addition to this point, a lower bound on $\langle q\bar q\rangle$ given in Eq.~(\ref{eq:bound on qqbar}) has not been reported in the previous literature.

\subsection{Our assumptions}
Before going to the main part, let us summarize our setup and assumptions in this paper.
The Lagrangian of QCD-like theory is given as
\begin{align}
{\cal L}_{\rm QCD} = -\frac{1}{4} G_{\mu\nu} G^{\mu\nu} + \sum_i \bar q_i (i\slashed{D}-m) q_i.
\end{align}
We take $\theta = 0$ and the quark mass $m$ to be non-negative. We assume $SU(N_f)$ invariance of the quark mass term for convenience of the analysis.
In addition, we take the following two assumptions:
\begin{enumerate}
\item In QCD-like theories with $\theta = 0$\footnote{
In the case of nonzero $\theta$, a non-trivial phase structure has been discussed in a lot of literature \cite{Dashen:1970et, Witten:1980sp, DiVecchia:1980yfw, Creutz:1995wf, Creutz:2003xu, Gaiotto:2017tne}.} and sufficiently small non-negative quark mass $m$ and $1/N_c$, there is no phase transition and we can take a smooth limit of $m\to 0$ and $1/N_c \to 0$.
\item The topological susceptibility in pure $SU(N_c)$ Yang-Mills theory and QCD-like theories with massive quarks is nonzero positive\footnote{
Note that Vafa-Witten theorem \cite{Vafa:1984xg} guarantees the topological susceptibility cannot be negative in pure Yang-Mills theory.}.
\end{enumerate}
Note that the assumption 1 leads that two limits $1/N_c \to 0$ and $m\to 0$ are commutable when $m$ is real positive.

\section{$N_f$ flavors of massless quarks}\label{sec:massless}
First, we review on Witten-Veneziano relation \cite{Witten:1979vv, Veneziano:1979ec}.
We assume $m=0$ in this section.
Let us define the following two-point correlation function:
\begin{widetext}
\begin{align}
\chi_t(p^2) \equiv -i\int d^4x e^{ipx} \langle0|
T
~\frac{g_s^2}{32\pi^2} G_{\mu\nu} \tilde G^{\mu\nu}(x)
~\frac{g_s^2}{32\pi^2} G_{\mu\nu} \tilde G^{\mu\nu}(0)
~|0\rangle. \label{eq:chi definition}
\end{align}
$\chi_t(0)$ is called as the topological susceptibility. Note that $\chi_t(0)$ is also obtained as $d^2 V/d\theta^2$,
where $V(\theta)$ is the vacuum energy of $\theta$-vacuum \cite{Jackiw:1976pf, Callan:1976je}:
\begin{align}
V(\theta) \equiv i \log \int {\cal D}A_\mu^a {\cal D}q {\cal D}\bar q \exp\left( i \int d^4x \left( {\cal L}_{\rm QCD} + \frac{\theta g_s^2}{32\pi^2} G_{\mu\nu} \tilde G^{\mu\nu} \right) \right).
\end{align}
\end{widetext}
When we have a massless quark, $\theta$ parameter can be absorbed by chiral rotation of quarks and $V(\theta)$ becomes independent on $\theta$, \textit{i.e.}, $\chi_t(0)=0$.

The assumption 1 allows us to use $1/N_c$ expansion to evaluate $\chi_t(p^2)$ by $1/N_c$ expansion:
\begin{align} 
\chi_t(p^2) = \chi_{t,g}(p^2) + \frac{N_f}{N_c} \chi_{t,q}(p^2) + {\cal O}(N_c^{-2}). \label{eq:chi large N}
\end{align}
$\chi_{t,g}(p^2)$ is the contribution from pure gluonic diagrams, which is the same as pure $SU(N_c)$ Yang-Mills theory.
$\chi_{t,g}(0)$ is nonzero because of the assumptions 2.
This term behaves as $N_c^0$ for large $N_c$ \cite{Witten:1979vv, Veneziano:1979ec, Witten:1998uka}.
$\chi_{t,q}(p^2)$ is the leading contribution from diagrams with single quark loop.
The leading contribution of quark loop is proportional to the number of quarks $N_f$ and suppressed by $1/N_c$ compared to $\chi_{t,g}(q^2)$.
We explicitly write $N_f/N_c$ factor in Eq.~(\ref{eq:chi large N}) so that $\chi_{t,q}(p^2)$ itself does not have $N_f$ and $N_c$ dependence at the leading term of $1/N_c$ expansion.

$\chi_t(0)$ should become zero once we introduce a massless quark \cite{Jackiw:1976pf, Callan:1976je}.
This means $\chi_{t,q}(0)$ term should cancel $\chi_{t,g}(0)$,
however, one could naively think this is impossible because of $N_f/N_c$ factor in front of $\chi_{t,q}$.
This puzzle has been solved by Witten \cite{Witten:1979vv} and Veneziano \cite{Veneziano:1979ec}
by assuming $\chi_{t,q}(p^2)$ has a pole from CP-odd scalar particle $\eta'$ whose mass squared scales as $1/N_c$.
Then, $\chi_{t,q}$ can be written as
\begin{align}
\chi_{t,q}(p^2) = \frac{a_{\eta'}^2}{p^2-m_{\eta'}^{(0)2}}.
\end{align}
where
\begin{align}
a_{\eta'} &\equiv \sqrt{ \frac{N_c}{N_f} } \times \langle 0| \frac{g_s^2}{32\pi^2} G_{\mu\nu} \tilde G_{\mu\nu} |\eta'\rangle. \label{eq:definition aeta}
\end{align}
$a_{\eta'}$ is defined so that $a_{\eta'}$ does not depend on $N_c$ and $N_f$ at the leading order of $1/N_c$ expansion.
$\chi_t(0)=0$ leads the following mass formula \cite{Witten:1979vv, Veneziano:1979ec}:
\begin{align}
m_{\eta'}^{(0)2} &= \frac{N_f}{N_c} \frac{a_{\eta'}^2}{\chi_{t,g}}. \label{eq:eta mass}
\end{align}
Here we attach the superscript to emphasize this is the $\eta'$ mass formula for $m=0$.
The chiral anomaly equation is
\begin{align}
\partial_\mu j_5^\mu = \frac{N_f g_s^2}{16\pi^2} G_{\mu\nu} \tilde G^{\mu\nu}.
\end{align}
Since $\langle 0 | G_{\mu\nu} \tilde G^{\mu\nu}  |\eta' \rangle$ is nonzero,  we obtain
\begin{align}
\langle 0 | j_5^\mu | \eta' \rangle
=
\sqrt{N_f} f p_\mu,
\quad f\equiv \frac{\sqrt{N_f \chi_{t,g}}}{2m_{\eta'}^{(0)}}.
\label{eq:decay constant eta}
\end{align}
This indicates $\eta'$ is the NG boson from spontaneous $U(1)_A$ symmetry breaking in the limit of $1/N_c \to 0$.
For finite $N_c$, $U(1)_A$ is explicitly broken by the chiral anomaly and $\eta'$ mass is nonzero and scales as $1/\sqrt{N_c}$.
In this definition, $f$ scales as $\sqrt{N_c}$ and independent on $N_f$ at the leading order of $1/N_c$.

Note that, even if $\eta'$ is identified as the NG boson for $U(1)_A$ symmetry breaking,
we do not know which order parameter induces this $U(1)_A$ symmetry breaking at this point.
For example, if the order parameter is 't Hooft determinant $\epsilon_{i_1 \dots i_{N_f}} \epsilon_{j_1 \dots j_{N_f}} (\bar q_{i_1} q_{j_1}) \cdots (\bar q_{i_{N_f}} q_{j_{N_f}})$, $U(1)_A$ is broken but $SU(N_f)_L \times SU(N_f)_R$ is unbroken \cite{Dvali:2017mpy}.

\section{$N_f$ flavors of quarks with small mass}\label{sec:massive}
Next, let us assume $m$ is nonzero positive. Nonzero quark mass breaks chiral symmetry explicitly and the remaining global symmetry is $SU(N_f)_V \times U(1)_V$.
We can derive the following Ward-Takahashi identities \cite{Crewther:1977ce, Crewther:1978kq, Shifman:1979if}
(for derivation, see the Appendix \ref{sec:WT identity}):
\begin{align}
N_f \chi_t(0) &= m \langle \bar q q\rangle + m^2 \chi_{s}(0) \label{eq:WT singlet}, \\
0 &= m \langle \bar q q\rangle + m^2 \chi_{adj} (0). \label{eq:WT adjoint}
\end{align}
where $\chi_{s}$ and $\chi_{adj}$ are pseudo-scalar susceptibility, which are defined as
\begin{align}
\chi_{s}(p^2) \equiv \frac{1}{N_f} \sum_{i,j} \int d^4x e^{ipx} \langle 0 | T (\bar q_i \gamma^5 q_i)(x) (\bar q_j \gamma^5 q_j)(0) | 0 \rangle, \label{eq:chi s}\\
\chi_{adj}(p^2) \delta_{ab} \equiv  \int d^4x e^{ipx} \langle 0 | T (\bar q_i \gamma^5 T^a q_i)(x) (\bar q_j \gamma^5 T^b q_j)(0) | 0 \rangle, \label{eq:chi adj}
\end{align}
Here $T^a$ are Hermite matrices such that ${\rm tr}[T^a T^b] = \delta_{ab}$ and ${\rm tr}T^a = 0$.
We have used the fact that $SU(N_f)_V \times U(1)_V$ symmetry is unbroken in the vacuum \cite{Vafa:1983tf} and parameterize $\langle \bar q_i q_j \rangle = \langle \bar q q \rangle \delta_{ij}$.

The assumption 1 implies the existence of $\eta'$ for at least sufficiently small $m$.
Because of the assumption 2, the cancellation between $\chi_{t,g}(0)$ and $\chi_{t,q}(0)$ in Eq.~(\ref{eq:chi large N}) should be broken for $m\neq 0$.
This means $\eta'$ mass formula should be modified as
\begin{align}
m_{\eta'}^2 = m_{\eta'}^{(0)2} + \delta m_{\eta'}^2.
\end{align}
$\delta m_{\eta'}^2$ is the leading contribution from the nonzero quark mass.
$\delta m_{\eta'}^2(m=0)=0$ is satisfied
and $\delta m_{\eta'}^2$ is an increasing function for quark mass $m$ at least if $m$ is sufficiently small.
The quark mass term is a source of the explicit breaking of $U(1)_A$ symmetry and this effect should remains in the limit of $1/N_c\to 0$.
Thus, $\delta m_{\eta'}^2$ scales as $N_c^0$ with given quark mass $m$.

\subsection{Singlet pseudo-scalar susceptibility}
In the current discussion, we have two small parameters; $m$ and $1/N_c$. Since $\eta'$ mass becomes small in the limit of small $m$ and $1/N_c$, some amplitude and correlation function could have singular behavior if there is $\eta'$ contribution. Let us discuss $\chi_s$ and $\chi_t$ with this point in mind.

First let us discuss $\chi_s$.
As we have seen, the leading contribution of quark loop in $\chi_t$ is coming from $\eta'$ one particle state.
Since $G_{\mu\nu} \tilde G^{\mu\nu}$ and $\sum_i \bar q_i \gamma^5 q_i$ have the same quantum number,
$\eta'$ gives dominant contribution to $\chi_s$ as
\begin{align}
\chi_{s}(k^2) \simeq \frac{1}{N_f} \frac{1}{k^2 - m_{\eta'}^{(0)2} - \delta m_{\eta'}^2} \left| \sum_i \langle 0 | \bar q_i \gamma^5 q_i| \eta' \rangle \right|^2.\label{eq:chi current albegra2}
\end{align}
For sufficiently small $m$, $\delta m_{\eta'}^2$ satisfies
\begin{align}
\delta m_{\eta'}^2 \ll M^2, \label{eq:parameter for chpt}
\end{align}
where $M$ is the typical mass of heavier hadrons.
Furthermore, since $\delta m_{\eta'}^2$ scales as $N_c^0$, 
we can take sufficiently small $1/N_c$ for given quark mass $m$ such that
\begin{align}
m_{\eta'}^{(0)2} \ll \delta m_{\eta'}^2. \label{eq:parameter for 1/nc expansion}
\end{align}
By using the assumptions 1, we assume that 
there is no phase transition between 
$m_{\eta'}^{(0)2} \ll \delta m_{\eta'}^2$ and
$m_{\eta'}^{(0)2} \gg \delta m_{\eta'}^2$
as long as both of $\delta m_{\eta'}^2$ and $m_{\eta'}^{(0)2}$ are sufficiently smaller than $M^2$.\footnote{
It is known that this assumption is broken in the case of $\theta = \pi$
\cite{Dashen:1970et, Witten:1980sp, DiVecchia:1980yfw, Creutz:1995wf, Creutz:2003xu, Gaiotto:2017tne}.
In this paper, we only focus on the case of $\theta = 0$ and take the asummption 1.
}
Then, in the case of $m_{\eta'}^{(0)2} \ll \delta m_{\eta'}^2 \ll M^2$, Eq.~(\ref{eq:chi current albegra2}) can be expanded as \footnote{
On the other hand, in the limit of $\delta m_{\eta'}^2 \ll m_{\eta'}^{(0)2} \ll M^2$, we obtain
\begin{align}
\chi_s(0) \simeq 
& -\frac{1}{N_f} \frac{1}{m_{\eta'}^{(0)2}} \left| \sum_i \langle 0 | \bar q_i \gamma^5 q_i| \eta' \rangle \right|^2 \nonumber\\
& +\frac{1}{N_f} \frac{\delta m_{\eta'}^2}{m_{\eta'}^{(0)4}} \left| \sum_i \langle 0 | \bar q_i \gamma^5 q_i| \eta' \rangle \right|^2.
\label{eq:chis 1}
\end{align}
Note that Eq.~(\ref{eq:chis 1}) cannot be directly derived from Eq.~(\ref{eq:chis 2}) and vice versa.
In this sense, we have to be careful about the order of taking limits.
See also Ref.~\cite{Grunberg:1985yg}.
}
\begin{align}
\chi_s(0) \simeq 
& -\frac{1}{N_f} \frac{1}{\delta m_{\eta'}^2} \left| \sum_i \langle 0 | \bar q_i \gamma^5 q_i| \eta' \rangle \right|^2 \nonumber\\
& +\frac{1}{N_f} \frac{m_{\eta'}^{(0)2}}{(\delta m_{\eta'}^2)^2} \left| \sum_i \langle 0 | \bar q_i \gamma^5 q_i| \eta' \rangle \right|^2.
\label{eq:chis 2}
\end{align}

Next, let us discuss $\chi_t(0)$.
In the case of $m_{\eta'}^{(0)2} \ll \delta m_{\eta'}^2 \ll M^2$,
by using Eq.~(\ref{eq:WT singlet}) and Eq.~(\ref{eq:chis 2}), we obtain
\begin{align}
N_f \chi_t(0)
\simeq
&~m \langle \bar q q \rangle
 -\frac{m^2}{N_f} \frac{1}{\delta m_{\eta'}^2} \left| \sum_i \langle 0 | \bar q_i \gamma^5 q_i| \eta' \rangle \right|^2 \nonumber\\
& +\frac{m^2}{N_f} \frac{m_{\eta'}^{(0)2}}{(\delta m_{\eta'}^2)^2} \left| \sum_i \langle 0 | \bar q_i \gamma^5 q_i| \eta' \rangle \right|^2.
\label{eq:chi current albegra3}
\end{align}
Let us compare the above equation with Eq.~(\ref{eq:chi large N}).
Since $m_{\eta'}^{(0)2} / \delta m_{\eta'}^2$ suppresses $(N_f/N_c)\chi_{t,q}(0)$ in Eq.~(\ref{eq:chi large N}),
the dominant contribution in $\chi_t(0)$ is from $\chi_{t,g}(0)$ and $\chi_t(0)$ becomes independent on $m$.
The assumption 1 implies that $\langle \bar q q\rangle$ is not singular at small $m$, \textit{i.e.}, $m\langle \bar q q \rangle$ term depends on $m$.
In order for the RHS of Eq.~(\ref{eq:chi current albegra3}) to be independent on $m$,
the first term and second term should cancel because the third term is suppressed by $m_{\eta'}^{(0)} / \delta m_{\eta'}^2$.
Then, we obtain the following two equations:
\begin{align}
m \langle \bar q q \rangle &= \frac{m^2}{N_f} \frac{1}{\delta m_{\eta'}^2} \left| \sum_i \langle 0 | \bar q_i \gamma^5 q_i| \eta' \rangle \right|^2, \\
N_f \chi_{t,g} &= \frac{m^2}{N_f} \frac{m_{\eta'}^{(0)2}}{(\delta m_{\eta'}^2)^2} \left| \sum_i \langle 0 | \bar q_i \gamma^5 q_i| \eta' \rangle \right|^2,
\end{align}
These equations correspond to Eq.~(1a) and Eq.~(1c) in Ref.~\cite{Veneziano:1980xs}.
We can show that
\begin{align}
\langle \bar q q \rangle &= 4f^2 \frac{\delta m_{\eta'}^2}{m} \label{eq:qqbar}, \\
\langle 0 | \bar q_i \gamma_5 q_j | \eta'\rangle &= \delta_{ij} \times \frac{2if}{\sqrt{N_f}} \frac{\delta m_{\eta'}^2}{m}, \label{eq:eta gamma5}\\
\chi_{s}(k^2) &= -\frac{4f^2}{k^2 - m_{\eta'}^{(0)2} - \delta m_{\eta'}^2} \frac{(\delta m_{\eta'}^2)^2}{m^2}, \label{eq:singlet channel}
\end{align}
where $f$ is defined in Eq.~(\ref{eq:decay constant eta}).
Note that these equations should be valid at the leading order of $m$ and $1/N_c$ as long as both of $\delta m_{\eta'}^2$ and $m_{\eta'}^{(0)2}$ are sufficiently smaller than $M^2$
though we have derived these relations by assuming $m_{\eta'}^{(0)2} \ll \delta m_{\eta'}^2 \ll M^2$.

Eq.~(\ref{eq:qqbar}) shows $\langle \bar q q \rangle$ obtains nonzero VEV for small nonzero $m$ because of $\delta m_{\eta'}^2 \neq 0$.
However, it is non-trivial whether $\langle \bar q q \rangle \neq 0$ in the limit of $m\to 0$.
We discuss this point in Sec.~\ref{sec:qqbar}.

\subsection{Adjoint pseudo-scalar susceptibility}
In $1/N_c$ expansion, the leading contribution to $\chi_{s}$ and $\chi_{adj}$ are the same
because they come from the similar diagram with connected quark lines.
In particular, by comparing with Eq.~(\ref{eq:singlet channel}), the behavior in the limit of large $k^2 \gg m_{\eta'}^{(0)2}$,
we obtain
\begin{align}
\chi_{adj}(k^2) = -\frac{4f^2}{k^2} \frac{(\delta m_{\eta'}^2)^2 }{m^2} \times (1 + {\cal O}(N_c^{-1}) ).
\end{align}
Thus, the one particle state should dominates at the leading order of $1/N_c$ expansion.
Therefore, $\chi_{adj}(k^2)$ to be consistent with Eqs.~(\ref{eq:WT adjoint}, \ref{eq:qqbar}) is given as
\begin{align}
\chi_{adj}(k^2) = -\frac{4f^2}{k^2 - \delta m_{\eta'}^2} \frac{(\delta m_{\eta'}^2)^2}{m^2},
\end{align}
and there should exist CP-odd $SU(N_f)$ adjoint scalar particles $\pi$ such that
\begin{align}
\langle 0 | \bar q \gamma^5 T^a q | \pi^a \rangle =
2if \frac{\delta m_{\eta'}^2 }{m}, \qquad
m_\pi^2 = \delta m_{\eta'}^2. \label{eq:pion}
\end{align}
We can immediately show that
\begin{align}
\langle 0 | \bar q \gamma^5 \gamma^\mu T^a q | \pi^a \rangle = if p^\mu.
\label{eq:decay constant}
\end{align}

\section{Massless quark, again}\label{sec:qqbar}
Finally, let us discuss massless QCD again.
The assumption 1 allows us to take a limit of $m\to 0$ of the results in the previous section.

\subsection{Chiral symmetry breaking}
In the massless quark limit, $m_\pi$ becomes $0$ because of Eq.~(\ref{eq:pion}) and $\delta m_{\eta'}^2 \to 0$.
On the other hand, the matrix element Eq.~(\ref{eq:decay constant}) does not depend on the quark mass $m$
and it keeps nonzero value.
Thus, in massless large $N_c$ QCD, we conclude there exist massless scalar particles which can be created by axial current operator.
This means chiral symmetry breaking $SU(N_f)_L \times SU(N_f)_R \to SU(N_f)_V$,
and $\pi$'s are NG bosons from this symmetry breaking.
Now we can see Eq.~(\ref{eq:qqbar}) is nothing but Gell-Mann--Oakes--Renner relation \cite{Gell-Mann:1968hlm}.

\subsection{Bilinear condensate} \label{sec:qqbar nonzero}
We have shown the chiral symmetry breaking $SU(N_f)_L \times SU(N_f)_R \to SU(N_f)_V$,
however, we have not specified the order parameter for this symmetry breaking.
As shown in Eq.~(\ref{eq:qqbar}), $\langle \bar q q \rangle$ in massless quark limit depends on the quark mass dependence in $\delta m_{\eta'}^2$.
For example, $\langle \bar q q\rangle$ becomes zero
if $\delta m_{\eta'}^2$ behaves as $m^n$ with $n\geq 2$.
The possibility of chiral symmetry breaking with $\langle \bar q q \rangle = 0$\footnote{
This possibility was not realized and $m_\pi^2 \propto m$ was implicitly assumed in Ref.~\cite{Veneziano:1980xs}.
\label{footnote:stern phase}}
 was pointed out by Refs.~\cite{Stern:1997ri, Stern:1998dy},
and later,
Kogan-Kovner-Shifman~\cite{Kogan:1998zc} excluded this possibility by utilizing a QCD inequality\footnote{
Note that chiral symmetry breaking without bilinear condensate can occur in a different matter content.
See, \textit{e.g.}, Ref.~\cite{Yamaguchi:2018xse}.}.
One of the QCD inequalities \cite{Weingarten:1983uj, Kogan:1998zc} yields
\begin{align}
\langle (\bar q_i \gamma^5 q_j)(x) (\bar q_j \gamma^5 q_i)(0) \rangle
\geq
|\langle (\bar q_i \gamma^\mu \gamma^5 q_j)(x) (\bar q_j \gamma^\nu \gamma^5 q_i)(0) \rangle|,
\label{eq:weingarten inequality}
\end{align}
where $i\neq j$.
Note that this inequality is exact and should be hold for any $\mu$, $\nu$ and any $x$.
For the derivation of this inequality, see the Appendix \ref{sec:QCD ineq}.
For $x\ll m_{\pi}^{-1}$, the pion contributions on both sides are
\begin{align}
\langle (\bar q_i \gamma^5 q_j)(x) (\bar q_j \gamma^5 q_i)(0) \rangle
&\simeq \frac{ \langle \bar q q \rangle^2 }{4\pi^2 f^2 x^2}, \\
\langle (\bar q_i \gamma^\mu \gamma^5 q_j)(x) (\bar q_j \gamma^\nu \gamma^5 q_i)(0) \rangle
&\simeq \frac{f^2}{8\pi^2} \left( \frac{g^{\mu\nu}}{x^4} - \frac{4x^\mu x^\nu}{x^6} \right).
\end{align}
Let us denote $M$ as the mass of the next to lightest particle which couples to $\bar q_i \gamma^5 q_j$ or $\bar q_i \gamma^\mu \gamma^5 q_j$.
For $M^{-1} \lesssim x \ll m_{\pi}^{-1}$, the pion contribution dominates both correlation function.
By comparing the pion contributions for $M^{-1} \lesssim x \ll m_{\pi}^{-1}$,
$\langle \bar q q \rangle$ cannot be zero and we obtain a lower bound
\begin{align}
\langle \bar q q \rangle \gtrsim f^2 M. \label{eq:bound on qqbar}
\end{align}
Here we are sloppy about ${\cal O}(1)$ numerical factor.
The QCD Lagrangian with massless quarks has non-anomalous discrete $Z_{2N_f}$ symmetry which is a subgroup of $U(1)_A$,
and nonzero $\langle \bar q q \rangle$ means spontaneous breaking of this $Z_{2N_f}$ symmetry.
This conclusion is consistent with an anomaly matching discussion in Ref.~\cite{Tanizaki:2018wtg}.

\section*{Acknowledgements}
The author thanks Ryuichiro Kitano and Kazuya Yonekura for useful discussions.

\appendix
\section{Derivation of Ward-Takahashi identities}\label{sec:WT identity}
In this Appendix, we derive Eq.~(\ref{eq:WT singlet}) and Eq.~(\ref{eq:WT adjoint}).
Similar equations were first derived in Refs.~\cite{Crewther:1977ce, Crewther:1978kq}.
The equivalent equations were used in Ref.~\cite{Veneziano:1980xs} to show chiral symmetry breaking in large $N_c$ QCD,
and also used in Ref.~\cite{Shifman:1979if} in the context of the strong CP problem.

For nonzero quark mass $m$, the chiral anomaly equation is
\begin{align}
\partial_\mu j_{5T}^\mu - 2im \bar q \gamma_5 T q &= {\rm tr}T \times \frac{g_s^2}{16\pi^2} G_{\mu\nu} \tilde G^{\mu\nu}, \label{eq:chiral anomaly} 
\end{align}
where $T$ is $N_f \times N_f$ matrix and $j_{5T}^\mu = \bar q \gamma^5 \gamma^\mu T q$.
By using Eq.~(\ref{eq:chi definition}) and Eq.~(\ref{eq:chiral anomaly}), $\chi_t(0)$ can be rewritten as
\begin{align}
({\rm tr}T)^2 \chi_t(0)  &= \chi_{t,v} + \chi_{t,m}, 
\end{align}
where
\begin{widetext}
\begin{align}
\chi_{t,v} &\equiv -\frac{i}{4} \lim_{p\to 0} \int d^4x e^{ipx} \langle 0| \left( \partial_\mu j_{5T}^\mu \right)(x) \left( \partial_\mu j_{5T}^\mu - 4im \sum_i \bar q \gamma^5 T q \right)(0) |0\rangle \\
\chi_{t,m} &\equiv m^2 \lim_{p\to 0} \int d^4x e^{ipx} \langle 0| \left(\bar q \gamma^5 T q\right)(x) \left(\bar q \gamma^5 T q\right)(0) |0\rangle.
\end{align}
$\chi_{t,v}$ is simplified by using Eq.~(\ref{eq:chiral anomaly}) again and taking integration by parts as
\begin{align}
\chi_{t,v}
&= -\frac{i}{4} \lim_{p\to 0} \int d^4x e^{ipx} \langle 0| \left( \partial_\mu j_{5T}^\mu \right)(x) \left( {\rm tr}T \frac{g_s^2}{32\pi^2}G_{\mu\nu} \tilde G^{\mu\nu} - 2im \bar q \gamma^5 T q \right)(0) |0 \rangle, \nonumber\\
&= \frac{i}{4} \lim_{p\to 0} \int d^4 x e^{ipx} \delta (x^0) \langle 0| [j_{5T}^0(x), (-2im \bar q \gamma^5 T \bar q)(0)] |0\rangle \nonumber\\
&= m \sum_i \langle 0| \bar q T^2 q |0\rangle.
\end{align}
We obtain the following equation:
\begin{align}
({\rm tr}T)^2 \chi_t(0)
=& m \langle 0| \bar q T^2 q |0 \rangle 
  + m^2 \lim_{p\to 0} \int d^4x e^{ipx} \langle 0| T(\bar q\gamma^5 T q)(x) (\bar q \gamma^5 T q)(0) |0\rangle. \label{eq:chi current albegra}
\end{align}
\end{widetext}
By using $\chi_s$ and $\chi_{adj}$ defined in Eqs.~(\ref{eq:chi s}, \ref{eq:chi adj}),
we obtain the following simple equations:
\begin{align}
N_f \chi_t(0) &= m \langle \bar q q\rangle + m^2 \chi_{s}(0), \\
0 &= m \langle \bar q q\rangle + m^2 \chi_{adj} (0).
\end{align}
Here we have used $\langle \bar q_i q_j \rangle = \langle \bar q q \rangle \delta_{ij}$.

\section{Derivation of QCD inequality} \label{sec:QCD ineq}
In this Appendix, we derive a QCD inequality Eq.~(\ref{eq:weingarten inequality}).
This inequality was first derived in Ref.~\cite{Weingarten:1983uj}.
Kogan-Kovner-Shifman \cite{Kogan:1998zc} pointed out this inequality can be used to exclude a possibility of $\langle \bar q q \rangle = 0$.
See also section 11 of Ref.~\cite{Nussinov:1999sx}.

The inequality Eq.~(\ref{eq:weingarten inequality}) can be shown in Euclidean QCD on a lattice.
The action is 
\begin{align}
S = S_U + \sum_{x,y} \bar q(x) D(x,y) q(y).
\end{align}
$x$ and $y$ are the lattice cites and $S_U$ is the action for link (gauge) field $U$.
$D(x,y)$ is given as \cite{Wilson:1974sk, Wilson:1975id, wilson1977quarks}
\begin{align}
D(x,y)
=& (4+m_0 a) \delta_{x,y} \nonumber\\
 & - \frac{1}{2} \sum_\mu \biggl[ (1-\gamma_\mu) U(x,y) \delta_{y,x+\hat\mu} \nonumber\\
 & \qquad\qquad + (1+\gamma_\mu) U(x,y) \delta_{y,x-\hat\mu} \biggr].
\end{align}
$m_0$ is the bare quark mass, $a$ is the lattice spacing, and $\hat\mu$ is the unit vectors for four directions.
Correlation function of operators which are made from quark fields is
\begin{align}
 &~\left\langle \prod_i q_{i_1}(x) \bar q_{i_2}(y) \right\rangle \nonumber\\
=&~
\displaystyle\frac{\displaystyle\int {\cal D}A \left[ \prod_i D^{-1}_{i_1 i_2}(x,y) \right] (\det D)^{N_f} e^{-S_U}}
{\displaystyle\int {\cal D}A (\det D)^{N_f} e^{-S_U}}.
\end{align}
Let us show
\begin{align}
\langle (\bar q_i \gamma^5 q_j)(x) (\bar q_j \gamma^5 q_i)(0) \rangle
\geq
|\langle (\bar q_i \gamma^M \gamma^5 q_j)(x) (\bar q_j \gamma^N \gamma^5 q_i)(0) \rangle|. \label{eq:CS inequality}
\end{align}
for any $M,N=0,1,2,3$ and $i\neq j$.

It is known that $\det D \geq 0$ is satisfied in QCD-like theories \cite{Vafa:1983tf}.
Then, a sufficient condition to satisfy the above inequality is
\begin{align}
&{\rm tr}[D^{-1}(x,y) \gamma^5 D^{-1}(y,x) \gamma^5] \nonumber\\
&\quad \geq \left| {\rm tr}[ D^{-1}(x,y) \gamma^5 \gamma^\mu D^{-1}(y,x) \gamma^5 \gamma^\nu] \right|.
\end{align}
for any $\mu,\nu=0,1,2,3$.
We can show $\gamma^5 D(x,y) \gamma^5 = D(y,x)^\dagger$,
then, $\gamma^5 D^{-1}(x,y) \gamma^5 = D^{-1}(y,x)^\dagger$.
The difference between LHS and RHS of the above inequality is
\begin{align}
 &{\rm tr}[D^{-1}(x,y) \gamma^5 D^{-1}(y,x) \gamma^5] \nonumber\\
 &\quad - \left| {\rm tr}[ D^{-1}(x,y) \gamma^5 \gamma^\mu D^{-1}(y,x) \gamma^5 \gamma^\nu] \right| \nonumber\\
=&{\rm tr}[D^{-1}(x,y) (D^{-1}(x,y))^\dagger ] \nonumber\\
 &\quad - \left| {\rm tr}[ D^{-1}(x,y) \gamma^\mu (D^{-1}(x,y))^\dagger \gamma^\nu] \right|. \label{eq:CS diff}
\end{align}
To show that this is non-negative, let us evaluate this expression explicitly.
Dirac matrices in four-dimensional Euclidean space are given as
\begin{align}
\gamma^0 = \left(\begin{array}{cc}
0 & 1 \\
1 & 0 
\end{array}\right), \quad
\gamma^i = \left(\begin{array}{cc}
0 & i\sigma^i \\
-i\sigma^i & 0 
\end{array}\right), \quad
\gamma^5 = \left(\begin{array}{cc}
1 & 0 \\
0 & -1 
\end{array}\right),
\end{align}
where $\sigma^i$ are the Pauli matrices.
Let us take the following $4 \times 4$ matrix $D^{-1}$:
\begin{align}
D^{-1} = \left(\begin{array}{ccc}
a_{11} & \cdots & a_{14} \\
\vdots & \ddots & \vdots \\
a_{41} & \cdots & a_{44} 
\end{array}\right).
\end{align}
An explicit calculation shows
\begin{align}
{\rm tr}[D^{-1} D^{-1\dagger}] =& \sum_{i,j} |a_{ij}|^2, \\
{\rm tr}[D^{-1} \gamma^0 D^{-1\dagger} \gamma^0 ] =& 2{\rm Re}( a_{11} a_{33}^* + a_{12} a_{34}^* + a_{13} a_{31}^* + a_{14} a_{32}^* \nonumber\\
&\quad + a_{21} a_{43}^* + a_{22} a_{44}^* + a_{23} a_{41}^* + a_{24} a_{42}^* ) \\
{\rm tr}[D^{-1} \gamma^3 D^{-1\dagger} \gamma^0 ] =& 2i{\rm Im}( -a_{11} a_{33}^* - a_{12} a_{34}^* - a_{13} a_{31}^* - a_{14} a_{32}^* \nonumber\\
&\quad + a_{21} a_{43}^* + a_{22} a_{44}^* + a_{23} a_{41}^* + a_{24} a_{42}^* )
\end{align}
By using Cauchy-Schwartz inequality, we can show
\begin{align}
{\rm tr}[D^{-1} D^{-1\dagger}] &\geq \left|{\rm tr}[D^{-1} \gamma^0 D^{-1\dagger} \gamma^0 ]\right|, \\
{\rm tr}[D^{-1} D^{-1\dagger}] &\geq \left|{\rm tr}[D^{-1} \gamma^3 D^{-1\dagger} \gamma^0 ]\right|.
\end{align}
In a similar way, we can show that Eq.~(\ref{eq:CS diff}) is non-negative for any $\mu$ and $\nu$,
and then, Eq.~(\ref{eq:CS inequality}) is satisfied.

\bibliography{ref}
\bibliographystyle{utphys}
\end{document}